\documentclass[preprint,preprintnumbers,amsmath,amssymb]{revtex4}
\usepackage[colorlinks=true, pdfstartview=FitV, linkcolor=blue, citecolor=red, urlcolor=magenta]{hyperref}
\usepackage{amssymb}
\usepackage{amsmath}
\usepackage{graphicx}
\usepackage{hyperref}
\usepackage{latexsym}
\usepackage{epsfig}
\usepackage{color}
\usepackage{comment}
\allowdisplaybreaks

\begin{document}

\title{\bf Probing Primordial Cosmology Through BBN Observational Constraints Under Extended Gravitational Dynamics}

\author{\textbf{$^{a,b}$Abdul Malik Sultan} \footnote{ams@uo.edu.pk, maliksultan23@gmail.com},\textbf{$^{a}$Manahil Ali} \footnote{manomanahil600@gmail.com}, \textbf{$^{c}$Muhammad Israr Aslam} \footnote{mrisraraslam@gmail.com, israr.aslam@umt.edu.pk} \textbf{$^d$Nazek Alessa} \footnote{naalessa@pnu.edu.sa}} \address{$^a$Department of Mathematics, University of Okara,  Okara-56300 Pakistan.\\$^b$Center for Theoretical Physics, Khazar University, 41 Mehseti Street, Baku, AZ1096, Azerbaijan\\ $^c$Department of Mathematics, School of Science, University of Management and Technology, Lahore-$54770$, Pakistan.\\$^d$Department of Mathematical Sciences, College of Science, Princess Nourah bint Abdulrahman University, P.O. Box 84428, Riyadh 11671, Saudi Arabia}

\begin{abstract}
In this article, We investigate the cosmological consequences of a recently developed $f(R,G,\mathcal{T})$ gravitational framework, in which the action is formulated as a general function of the Ricci scalar $R$, the Gauss-Bonnet invariant $G$, and the trace of the energy-momentum tensor $\mathcal{T}$. As one of the most reliable probes of physical conditions in the early Universe, BBN offers a stringent framework for testing deviations from standard cosmology. We consider four representative models and constrain them using observational limits on $\left|\Delta T_f/T_f\right|$ and the primordial helium mass fraction $Y_p$. The bounds obtained identify the allowed parameter regions for each model and demonstrate that significant departures from standard cosmology are compatible with nucleosynthesis observations. Our analysis shows that broad regions of the parameter space satisfy existing nucleosynthesis constraints, indicating the consistency of $f(R,G,\mathcal{T})$ gravity with the observed primordial light-element abundances and the established picture of the early Universe preserving the observed abundances of light nuclei.
\end{abstract}
\maketitle

\section{Introduction}

The $\Lambda$CDM model, often referred to as the standard cosmological paradigm, successfully describes the evolution of the Universe from the Big Bang to its current accelerated expansion within the framework of General Relativity (GR). Despite its empirical success, the $\Lambda$CDM model depends on dark energy (DE) and dark matter (DM) of unknown origin. 
Observational evidence from Type Ia supernovae ($SnIa$), the cosmic microwave background (CMB), and large-scale structure (LSS) strongly indicates that the current accelerated expansion of the Universe is driven by DE \cite{2}-\cite{4}. Motivated by these issues, modified theories of gravity have been widely explored as viable alternatives to DE.  In this context, $f(R,G,\mathcal{T})$ gravity extends GR by incorporating curvature, Gauss-Bonnet, and matter contributions, offering a unified geometric framework to explain cosmic acceleration without invoking unknown energy components \cite{6}.

Big Bang nucleosynthesis (BBN), which occurred within the first few minutes after the Big Bang as the universe cooled sufficiently for nuclear reactions to take place, resulted in the formation of the lightest stable nuclei, mainly hydrogen and helium $(He)$, along with small amounts of deuterium, helium-3 $(^3He)$ and lithium-7 $(^7Li)$ \cite{7}. Since its early development, BBN has remained a cornerstone of early Universe studies. At the same time, GR has provided the theoretical framework for describing gravity and governing cosmic expansion during this crucial epoch. Within the framework of GR, the standard BBN scenario has successfully explained the primordial abundances of light nuclei, including deuterium, $(^3He)$, $(^4He)$, and $(^7Li)$ \cite{9}.

Although General Relativity (GR), formulated within the framework of Riemannian geometry, provides the theoretical foundation of modern cosmology \cite{10,11}, several unresolved cosmological problems have motivated the study of extended theories of gravity. The theoretical foundations of BBN were established through the pioneering works of Gamow, Alpher, and Herman, and subsequent developments have made BBN one of the most successful probes of the early Universe \cite{12,13,14}. Since the primordial abundances of light elements depend sensitively on the expansion rate of the Universe, even small modifications to gravitational dynamics can affect the freeze-out process and the resulting elemental abundances. Therefore, BBN provides a valuable observational framework for testing extended theories of gravity and constraining possible deviations from the standard cosmological scenario.

In the broader framework of cosmological research, the primary objective is to construct a consistent picture of the universe by linking fundamental theory with high precision observations. early Universe phenomena are particularly important in this respect, since even small deviations from standard assumptions can leave observable imprints on later cosmic evolution. Processes governed by the expansion history, therefore, provide a powerful testing ground for gravity under extreme conditions. This perspective has gained further significance due to ongoing theoretical challenges and observational tensions that remain unresolved within conventional models. As a result, modern cosmology has evolved into a key framework not only to describe the history of the universe but also to critically examine and constrain extensions of gravitational theory \cite{13}.

Observational puzzles such as the accelerated expansion of the universe, the unknown nature of DM, and the mechanisms driving inflation have raised questions about the completeness of the Einstein–Hilbert formulation. In response, a wide class of modified gravity theories has been proposed in which the gravitational action is enriched by additional geometric ingredients or new dynamical degrees of freedom. 
 These theories extend the Einstein-Hilbert action through modifications involving curvature, torsion, higher-order geometric terms, scalar fields, or non-metricity. Representative examples include $f(R)$ gravity \cite{14}, $(f(T)$ gravity \cite{15}, Gauss-Bonnet and Lovelock theories \cite{6}, scalar-tensor models \cite{17}, and $(f(Q)$ gravity \cite{18}. Their cosmological implications have been widely investigated, particularly regarding cosmic expansion, structure formation, and early Universe evolution \cite{21}-\cite{28}.

Several investigations have explored the role of alternative cosmological scenarios in shaping the dynamics of BBN, emphasizing how modifications to the early Universe expansion rate can affect the freeze-out process and the primordial abundances of light elements. 
 BBN constraints have been investigated in a variety of modified gravity frameworks, including $f(T)$ gravity \cite{29}, $f(R,T)$ gravity \cite{30}, entropy-modified cosmologies \cite{31}, $f(T,T_G)$ gravity \cite{32}, higher-order and Gauss-Bonnet extensions \cite{33}, $f(Q,T)$ gravity \cite{34}, as well as $f(Q)$ gravity \cite{35}. In addition, models with a time-dependent gravitational coupling have also been constrained using primordial element abundances \cite{36}. More recent investigations have extended BBN tests to Weyl-type $f(Q,T)$ gravity \cite{37}, primordial black hole (PBHs) scenarios \cite{38}, and other extended gravity frameworks, while combined analyses with BBN and complementary cosmological observations have provided increasingly precise constraints on fundamental parameters \cite{39}. Further studies have also examined extended teleparallel gravity \cite{40} and energy-momentum-squared gravity \cite{41}. Recent studies have also explored the connection between early and late-time cosmological observations. In particular, combined analyses involving BBN, cosmic chronometers (CC),  Planck, Baryon Acoustic Oscillations (BAO), and other observational datasets have provided complementary constraints on modified gravity models and their parameters \cite{42,43}. Additional investigations have examined the effects of boundary corrections, entropy modifications, dark matter scenarios, generalized uncertainty principle corrections, and several extended gravity frameworks on the BBN epoch \cite{44,45,46,47,47a,47b,mali}. These studies further demonstrate the broad applicability of primordial nucleosynthesis as a powerful probe of non-standard gravitational and cosmological scenarios. Collectively, these studies show that even small deviations from the standard expansion history can leave observable imprints on BBN, providing useful bounds on the parameters of extended gravitational theories.

 Although BBN provides important constraints on the gravitational dynamics of the early Universe, a complete assessment of modified gravity models also requires considering their behavior at later cosmological epochs. Recent studies have therefore investigated early- and late-time observations together in different modified gravity frameworks. For example, BBN constraints have been combined with CC and gamma-ray burst observations in \(f(T)\) gravity, while \(f(Q)\) gravity has been tested using BBN alongside CC, Pantheon+SH0ES and BAO data. Similar analyses have been carried out in \(f(T,\mathcal{T})\) and \(f(T,\mathcal{L}_m)\) gravity using both primordial nucleosynthesis and late-time observations \cite{47c,47d,47e,47f}. These studies highlight the complementary nature of early- and late-time probes, as BBN is sensitive to changes in the expansion history during the primordial era, whereas late-time observations constrain the subsequent cosmic evolution. In this context, the present work establishes BBN constraints on the considered \(f(R,G,\mathcal{T})\) models, while a detailed investigation using late-time cosmological observations is left for future work.

Although BBN has been widely employed to constrain modified gravity theories, the combined role of curvature, higher-order geometric, and matter contributions in \(f(R,G,\mathcal{T})\) gravity provides a distinct motivation for studying this framework during the early Universe. In this theory, simultaneous dependence on \(R\),  \(G\), and \(\mathcal{T}\) can modify the history of cosmological expansion and consequently affect the dynamics of BBN. Since the expansion rate directly influences the weak-interaction freeze-out and the subsequent production of light elements, BBN offers a sensitive test of such gravitational modifications. The central hypothesis of this work is that the considered \(f(R,G,\mathcal{T})\) models may admit viable parameter regions in which the modified expansion history remains consistent with BBN observations. We therefore examine their effects on the freeze-out temperature and primordial helium abundance and use the resulting constraints to identify the viable parameter regions of the models.


The structure of this manuscript is as follows. In Section II, we present the basic formulation of
$f(R,G,\mathcal{T})$ gravity theory and derive the corresponding gravitational field equations. Section III is devoted to the cosmological setup relevant to the early Universe, where we establish the modified Friedmann equations and implement BBN constraints. In Section IV, we confront these theoretical predictions with observational bounds by analyzing a representative model $f(R,G,\mathcal{T})$ and examining the allowed parameter space. Section V analyzes the behavior of the $He$ mass fraction $Y_p$ for four different gravity models. Finally, Section VI summarizes the principal results and discusses their significance for the consistency and applicability of $f(R,G,\mathcal{T})$ gravity in the context of early Universe cosmology.

  \section{ $f(R, G, \mathcal{T})$ GRAVITY Overview and field equations}

In recent years, several generalizations of Einstein’s theory of GR have been developed to explore possible deviations in gravitational physics at cosmological scales. Inspired by these advancements, Debnath \cite{52} generalized the gravitational action by proposing the $f(R,G,\mathcal{T})$ theory, where the Lagrangian involves simultaneous dependence on $R$, $G$ and $\mathcal{T}$. He demonstrated that a more comprehensive model can satisfy standard energy conditions and remain stable under a power-law expansion scenario. Additionally, in \cite{53}, the authors conducted a detailed investigation of cosmographic behavior and a variety of cosmological tests within the framework $f(R,G,\mathcal{T})$ , establishing important criteria for the viability and consistency of the theory with observational data. The action term corresponding to this generalized $f(R,G,\mathcal{T})$ gravity model can be given in the following form \cite{52,53}
\begin{eqnarray}\label{1}
S= \int \sqrt{-g} d^4x \bigg[\frac{1}{2 \kappa^2} f(R,G,\mathcal{T})+\mathbf{L_m}\bigg], 
\end{eqnarray}
where $f(R,G,\mathcal{T})$ denotes a general function, the term $\mathbf{L_m}$ corresponds to the Lagrangian density of matter, while $g=|g_{\alpha \beta}|$ and $\kappa^2=8 \pi G_N$. Throughout the analysis, we adopt the commonly used convention $8\pi G_N=C=1$ to simplify the equations, where $G_N$ represents Newton’s gravitational constant. 
The energy-momentum tensor of the matter field is expressed as follows \cite{53aa}
\begin{eqnarray}\label{1a}
T_{\alpha \beta}=\frac{-2}{\sqrt{-g}} \frac{\delta(\sqrt{-g}\mathbf{L_m})}{\delta g^{\alpha \beta}}.   
\end{eqnarray}
The trace of the energy-momentum tensor is defined by {\color{red}\(\mathcal{T} = g^{\alpha \beta}T_{\alpha \beta}\)}. Assuming that the Lagrangian matter \(\mathbf{L}_m\) depends solely on the metric tensor \(g_{\alpha \beta}\), we obtain the following.
\begin{eqnarray}\label{1b}
 T_{\alpha \beta}=  g_{\alpha \beta} \mathbf{L}_m-2\frac{\partial\mathbf{L}_m}{\partial g^{\alpha \beta}}.
\end{eqnarray}
Varying the action given in Eq. (\ref{1}) with respect to the metric tensor yields the following result. variation
\begin{eqnarray}\label{1c}
\delta_s=\frac{1}{2 \kappa^2}\int\bigg[f_R \delta R+f_G \delta G+f_{\mathcal{T}}\delta \mathcal{T}+f\delta (\sqrt{-g})+\frac{2 \kappa^2}{\sqrt{-g}}\delta \left(\mathbf{L}_m (\sqrt{-g}\right)\bigg]\sqrt{-g}d^4x,    
\end{eqnarray}
here $f_R=\frac{\partial f}{\partial R},~f_G=\frac{\partial f}{\partial G}$ and $f_\mathcal{T}=\frac{\partial f}{\partial \mathcal{T}}$. The explicit forms of the curvature scalar $R$, the Gauss–Bonnet invariant $G$, and the trace $\mathcal{T}$ employed in this framework follow the definitions given in \cite{53}.
\begin{eqnarray}\label{2}
R= g^{\alpha \beta} R_{\alpha \beta}, \, G= R^2-4 R_{\alpha \beta} R^{\alpha \beta}+ R_{\alpha \beta \mu \nu} R^{\alpha \beta \mu \nu}, \, \mathcal{T}= g^{\alpha \beta} T_{\alpha \beta}. 
\end{eqnarray}
These terms account for curvature, higher-order geometric effects, and matter-geometry coupling, respectively; their combined contribution can alter the cosmic expansion rate and is therefore important for studying BBN.
Varying the action with respect to the metric tensor yields the field equations of \(f(R,G,\mathcal T)\) gravity, which can be written as \cite{53}
\begin{eqnarray}\nonumber
& &\left( R_{\alpha\beta} + g_{\alpha\beta} \nabla^{2} 
- \nabla_{\alpha}\nabla_{\beta} \right) f_{R}
+ \left(2 R R_{\alpha\beta} - 4 R_\beta^\mu R_{\alpha \, \mu}
- 4 R_{\alpha\mu\beta\nu} R^{\mu\nu}
+ 2 R_\alpha^{\ \mu \nu \delta} R_{\beta \mu\nu\delta}
\right) f_{G}\\ \nonumber  &-& \frac{1}{2} f\, g_{\alpha\beta} + \big(
2 R g_{\alpha\beta} \nabla^{2}
- 2 R \nabla_{\alpha}\nabla_{\beta}
- 4 g_{\alpha\beta} R^{\mu\nu} \nabla_{\mu}\nabla_{\nu}
- 4 R_{\alpha\beta} \nabla^{2}
+ 4 R^\mu_\alpha \nabla_{\beta}\nabla_{\mu}
+ 4 R^\mu_\beta \nabla_{\alpha}\nabla_{\mu}
\\\label{3} &+& 4 R_{\alpha \mu \beta \nu} \nabla^{\mu}\nabla^{\nu}
\big) f_{G} = T_{\alpha\beta}
- \left( T_{\alpha\beta} + \Theta_{\alpha\beta} \right) f_{\mathcal{T}}.    
\end{eqnarray}
In the above expressions, the quantities $f_R=\frac{\partial f}{\partial R},~f_G=\frac{\partial f}{\partial G},~f_\mathcal{T}=\frac{\partial f}{\partial \mathcal{T}}$ represent the partial derivatives of the function $f(R,G,\mathcal{T})$ with respect to its arguments. The operator $\nabla ^2= \nabla_\alpha \nabla^\alpha$ denotes the D’Alembertian operator acting on scalar functions. The tensor $T_{\alpha \beta}$ corresponds to the energy-momentum tensor associated with the matter distribution, and its trace is denoted by $\mathcal{T}$. For a perfect (ideal) fluid configuration, the energy-momentum tensor takes the following standard form:
\begin{eqnarray}\label{4}
T_{\alpha \beta}=(\rho+p) u_\alpha u_\beta +p g_{\alpha \beta}.   
\end{eqnarray}
Here, $\rho$ and $p$ denote the energy density and pressure of the perfect fluid, respectively. The four-velocity vector obeys the normalization condition $ u_\alpha u^\alpha=-1$, and its geodesic motion implies $ u^\alpha \nabla_\beta u_\alpha=0$. Furthermore, the tensor associated with the matter-geometry coupling is defined as $\Theta_{\alpha\beta} = -2\, T_{\alpha\beta} + p\, g_{\alpha\beta}$. For the cosmological background, we adopt the spatially flat Friedmann- Lemaître- Robertson- Walker (FLRW) universe, whose line element is expressed as follows:
\begin{eqnarray}\label{5}
ds^2=-dt^2 + a^2(t) \bigg( dr^2 +r^2(d\theta^2+sin^2\theta d\phi^2)\bigg).  
\end{eqnarray}
Here, $a(t)$ denotes the cosmological scale factor. Using Eq. (\ref{2}), we obtain the following set of relations:
\begin{eqnarray}\label{6}
 R=6(\dot{H}  +2H^2), \, G=24 H^2(\dot{H}+H^2). 
\end{eqnarray}
In this context, the Hubble parameter is defined as $H=\frac{\dot{a}}{a}$, where the overhead point denotes the differentiation with respect to cosmic time $t$.
Using the FLRW metric, the trace of the energy-momentum tensor becomes
\begin{eqnarray}\label{7}
\mathcal{T}=3p-\rho= (3 \omega-1)\rho,    
\end{eqnarray}
where $\omega$ denotes the equation of state (EoS) parameter.
Furthermore, for an ideal fluid, the conservation law $\nabla_\alpha T_\beta ^\alpha=0$ leads to the following relation:

\begin{eqnarray}\label{8}
\dot{\rho}+3H(\rho+p)=0.    
\end{eqnarray}
With these assumptions in place, the modified Friedmann equations derived from Eq. (\ref{3}) can be written in the following form:
\begin{eqnarray}\nonumber
3H^{2} &=& \frac{1}{f_{R}} \big[ 
\rho + (\rho + p)\, f_{\mathcal{T}}
+ \frac{1}{2}\big( R f_{R} - f \big)
- 3H \dot{f}_{R}
+ 12 H^{2}(\dot{H} + H^{2})\, f_{G} \\\label{9} &-&12 H^3\dot{f}_G
\big]. \\\nonumber
2\dot{H} + 3H^{2}&=& -\frac{1}{f_{R}} \big[p - \frac{1}{2}(R f_{R} - f)+ 2H \dot{f}_{R}+ \ddot{f}_{R}- 12H^{2}(\dot{H} + H^{2}) f_{G}+8H \\\label{10} &\times& (\dot{H} + H^{2}) \dot{f}_{G}+4H^{2}
\ddot{f}_{G}\big].   
\end{eqnarray}
The Friedmann equations derived above establish the general cosmological description within the $f(R,G,\mathcal{T})$ framework; their particular form is determined by the choice of the function $f(R,G,\mathcal{T})$.

The field equations derived above serve as the groundwork for studying possible consequences of the $f(R,G,\mathcal{T})$ gravity in the expansion of early Universe. Specifically, alterations in the expansion rate affect the weak processes that cause conversion of a neutron into a proton, influencing both the freeze-out temperature and production of light elements during BBN. This research studies how different forms of $f(R,G,\mathcal{T})$ may change the Universe’s early evolution and whether those changes agree with observations. The subsequent section discusses the specific models we study and why we chose them, as well as the impact of the models on BBN. Using the freeze-out temperature and primordial helium to constrain the model parameters using the main observables for BBN.

\section{The Standard BBN Picture: Conditions and Evolution}

In this section, we examine how variations in the freeze-out temperature $\mathbf{T}_f$ relate to the observational bounds imposed by BBN. Because BBN unfolds during the radiation-dominated era of cosmic evolution \cite{54,55}, we adopt the conventional radiation-driven cosmological model for our analysis. In this stage, the first Friedmann equation assumes a simplified form, which provides an appropriate baseline for evaluating the impact of modified gravity effects on the thermal history of the early Universe.
\begin{eqnarray}\label{11}
H^2 = \frac{\rho_r}{3M_P^2}\approx H^2_{GR},
\end{eqnarray}
Here, $M_p$ is the Planck mass defined as $M_p=\frac{1}{\sqrt{8\pi G}}$ and the quantity $ \rho_r$ denotes the energy density of relativistic species, which governs the thermal and dynamical behavior of the universe during the radiation-dominated phase. It can be expressed as
\begin{eqnarray}\label{13}
\rho_r=  \frac{\pi^2g_*}{30}\mathbf{T}^4.
\end{eqnarray}
In this calculation, $\mathbf{T}$ refers to the temperature and $g_*=g(\mathbf{T})$ represents the effective relativistic degrees of freedom that contribute to the total radiation energy density. For the cosmological conditions relevant to the BBN era, $g_*$ is commonly approximated to be around $10$. Substituting the expressions for the Planck mass $M_P$ and the radiation energy density $\rho_r$ into Eq. (\ref{11}) leads to a more compact form of the equation
\begin{eqnarray}\label{14}
H(\mathbf{T})= \bigg(\frac{4\pi^3g_*}{45}\bigg)^\frac{1}{2} \frac{\mathbf{T}^2}{M_{pl}}.
\end{eqnarray}
In this analysis, the reduced Planck mass is related to the conventional Planck mass through $M_{pl}=\sqrt{8\pi}M_P$. During the radiation-dominated stage of cosmic evolution, the conservation of radiation dictates that the scale factor grows as $a(t)\sim \sqrt{t}$. Consequently, the Hubble parameter acquires the familiar time dependence $H=\frac{1}{2t}$. Using this relation, one can connect the cosmic temperature to the expansion time via: 
\begin{eqnarray}\label{15}
\frac{1}{t}= \bigg(\frac{16\pi^3g_*}{45}\bigg)^{1/2} \frac{\mathbf{T}^2}{M_{pl}},   
\end{eqnarray}
 which may also be rearranged to give an approximate temperature-time relation of the form 
\begin{eqnarray}\nonumber
\mathbf{T}(t)\simeq (t/s)^{-1/2}MeV.     
\end{eqnarray}
During the BBN era, the neutron population is primarily generated and regulated through weak interaction processes that continually convert a fraction of protons into neutrons \cite{7}. The role of this process is that neutrons decay to protons through three main decay modes:
\begin{eqnarray}\nonumber
n\rightarrow p+e^-+\bar{v}_e \\\nonumber n+v_e\rightarrow p+e^- \\\nonumber n+e^+\rightarrow p+\bar{v}_e,
\end{eqnarray}
respectively.
The neutron abundance is therefore determined by the interplay of these weak reaction rates and the rapidly evolving thermal background.
\begin{eqnarray}\label{16}
\Lambda_{pn}(\mathbf{T})= \Lambda_({n+v_e \rightarrow p+e^-})+\Lambda_({n+e^+\rightarrow p+\bar{v}_e}) + \Lambda_({n\rightarrow p+e^- + \bar{v}_e}).
\end{eqnarray}
The reverse transition, converting neutrons back into protons, is characterized by the rate $\Lambda_{np}(\mathbf{T})$. Together, the forward and backward processes give the overall weak conversion rate as:
\begin{eqnarray}\label{17}
\Lambda_{tot}(\mathbf{T})= \Lambda_{pn}(\mathbf{T})+\Lambda_{np}(\mathbf{T}).    
\end{eqnarray}
Under the assumption that all interacting species share a common temperature low enough for classical statistics to provide a good approximation, the Boltzmann distribution may be used in place of the Fermi–Dirac form. In this regime, the electron mass can be neglected since the energies of electrons and neutrinos are typically much larger. These simplifications enable standard approximations that allow one to estimate the neutron fraction. By analyzing the dominant proton-to-neutron processes within this low-temperature limit, we arrive at a compact expression for the total reaction rate, as outlined in several foundational studies \cite{29,31,59,60}.
\begin{eqnarray}\label{18}
\Lambda_{tot}(\mathbf{T}) = 8(12\mathbf{T}^2+6\mathcal{Q} \mathbf{T}+\mathcal{Q} ^2)A\mathbf{T}^3,
\end{eqnarray}
in this formulation $\mathcal{Q} $ represents the neutron–proton mass gap, given by 
\begin{eqnarray}\nonumber
\mathcal{Q}&=&m_n-m_p= 1.29 \times 10^{-3}GeV, \\\nonumber  
 A&=&1.02 \times 10^{-11} MeV^{-4} 
\end{eqnarray}
The primordial $He$ fraction is written as $Y_p=\lambda \frac{2x(\mathbf{T}_f)}{1+x(\mathbf{T}_f)}$, where $\lambda = e^{({\mathbf{T}_f-\mathbf{T}_n}/\tau)}$. Here, $\mathbf{T}_f$ is the freezing temperature of the weak interaction, $\mathbf{T}_n$ marks the beginning of nucleosynthesis, and $\tau$ denotes the free-neutron lifetime, measured as $877.75 \pm 0.28$ seconds \cite{53a}. The equilibrium neutron-to-proton ratio at freeze-out is $x(\mathbf{T}_f) = e^{-\frac{\mathcal{Q}}{\mathbf{T}_f}}$.  
The function $\lambda(\mathbf{T}_f)$ quantifies the fraction of neutrons that decay between $\mathbf{T}_f$ and $\mathbf{T}_n$. To identify the freeze-out point, we equate the expansion timescale $H^{-1}$ with $\tau_{tot}(\mathbf{T})$. Thermal equilibrium holds as long as interactions proceed faster than cosmic expansion \cite{54,62}; decoupling occurs once the reverse becomes true. The freeze-out temperature satisfies
 \begin{eqnarray}\label{20}
H(\mathbf{T}_f) =\Lambda(\mathbf{T}_f) \simeq c_q \mathbf{T}_f^5 ,    
 \end{eqnarray}
with $c_q=96A \simeq 9.8 \times 10^{-10}$GeV$^{-4}$ \cite{29,31,59,60}. Using Eqs. (\ref{14}) and (\ref{17}), this criterion leads directly to the required expression.
\begin{eqnarray}\label{21}
\mathbf{T}_f= \bigg(\frac{4\pi^3g_*}{45c_q^2 M_{pl}^2}\bigg)^{1/6}.
\end{eqnarray}
Within modified cosmological scenarios, the Hubble expansion rate  $H$ no longer exactly matches its standard GR counterpart $H_{GR}$. This discrepancy introduces a correction in the freeze-out temperature, denoted by $\Delta$. This shift in the decoupling temperature naturally affects the predicted abundance of light elements. In particular, the primordial $He$ fraction 
$Y_p$ acquires a corresponding modification and can therefore be expressed as
\begin{eqnarray}\label{22}
\Delta Y_p= Y_p\bigg[\bigg(1-\frac{Y_p}{2\lambda}\bigg)\ln \bigg(\frac{2\lambda}{Y_p}-1\bigg)-\frac{2\mathbf{T}_f}{\tau}\bigg]\frac{\Delta \mathbf{T}_f}{\mathbf{T}_f}.
\end{eqnarray}
We impose the condition $\Delta \mathbf{T}(\mathbf{T}_n)=0$, since the temperature $\mathbf{T}_n$ is fixed by the deuterium binding energy \cite{63,64}. Observational analyses of the BBN era indicate that the mass fraction $He$ lies within the range reported in \cite{65}–\cite{70}.
\begin{eqnarray}\label{23}
Y_p= 0.245\pm0.003, \hspace{2cm} \left|\Delta Y_p \right|< 10^{-4}.
\end{eqnarray}
When this analysis is extended to a modified theory of gravity, the Friedmann equations acquire an extra contribution arising from the gravitational modification. To remain consistent with the well-established observational bounds from the BBN era, this additional term must remain much smaller than the usual radiation energy density. Using the generalized expression for the modified Friedmann equation, we obtain the condition 
$3M^2_pH^2 = \rho_m + \rho_r + \rho$, here $\rho_m$ denotes the energy density of matter. From this  equation, we can derive the relevant constraints as follows
\begin{eqnarray}\label{24}
H = H_{GR} \bigg(1+\frac{\rho}{\rho_r}\bigg)^{1/2} = \Delta H + H_{GR}.
\end{eqnarray}
Here, the quantity $ H_{GR}= \sqrt{\rho_r/3M_p^2}$ describes the expansion rate of the Universe in standard GR. This relation allows us to infer the corresponding dynamical evolution of the early cosmos and provides the reference point against which deviations from modified gravity models can be assessed.
\begin{eqnarray}\label{25}
\Delta H=H_{GR} \bigg(\sqrt {1+{\frac{\rho}{\rho_r}}}-1\bigg).
\end{eqnarray}
Because the evolution of the universe is non-standard, this also has an impact on the freeze-out temperature $\Delta \mathbf{T}_f$, which in turn affects the statistical splits. Take this equation together with Eq. (\ref{21}).
By merging this relation with the freeze-out condition, the condition for the weak interaction rate to be met at the freeze-out temperature is as follows: $\Lambda(\mathbf{T})=H(\mathbf{T}_f)$, where $\Lambda(\mathbf{T})=c_{q}\mathbf{T}_{f}^{5}$. An initial first-order expansion of this relationship around the freeze-out temperature is obtained.
\begin{eqnarray}\nonumber
\Delta H= \Delta \Lambda= \frac{d \Lambda}{d \mathbf{T}}\bigg|_{\mathbf{T}_f} \Delta \mathbf{T}_f.
\end{eqnarray}
Since $\Lambda(\mathbf{T})=c_{q}\mathbf{T}_{f}^{5}$, 
\begin{eqnarray}\nonumber
    \frac{d \Lambda}{d \mathbf{T}}=  5c_{q}\mathbf{T}_{f}^{4},
\end{eqnarray}
therefore,
\begin{eqnarray}\label{26}
\Delta H= 5c_q \mathbf{T}_f^4 \Delta \mathbf{T}_f.
\end{eqnarray}
In the regime where the contribution of DE, $\rho$, remains negligible compared to the radiation density $\rho_r$, the resulting expression can be reduced to a much simpler form.
\begin{eqnarray}\label{27}
\bigg|\frac{\Delta \mathbf{T}_f}{\mathbf{T}_f}\bigg| \simeq \frac{\rho}{\rho_r}\frac{H_{GR}}{10c_q \mathbf{T}_f^5}.
\end{eqnarray}
By matching the theoretically obtained relation with the latest observational bounds, we arrive at the corresponding numerical estimate as
\begin{eqnarray}\label{28}
\left| \frac{\Delta \mathbf{T}_f}{\mathbf{T}_f} \right| < 4.7 \times 10^{-4}.
\end{eqnarray}
This value is inferred from observational measurements of the baryonic mass fraction transformed into $^4He$, as given in Eq. (\ref{23}) \cite{65}–\cite{70}.

In the discussion that follows, we assess how modified gravity can impact the freeze-out period of BBN approximately. The freeze-out temperature is set by the limit between the rate at which the universe expands and the rate of weak interactions that can turn a neutron into a proton. If the expansion rate differs from what is expected according to standard GR, the freeze-out will happen at a later (or earlier) time, shifting the neutron-to-proton ratio and the production of the primordial helium. Here we systematically studied the effect of modified gravity using a first-order perturbative treatment of the freeze-out condition around standard GR. Such background bumps allow one to determine observational limits that can be placed on the additional gravitational contribution to the earliest universe through BBN constraints, as already done in this paper. But of course, this limitation arises because this analysis, as it is here presented, relies on this approximate description, not on a full numerical simulation of the BBN nuclear reaction network; therefore, this should come with some caveats concerning the limits of the assumptions and approximations that are made here.

\section{BBN in $f(R,G,\mathcal{T})$ Gravity}

In this section, we apply the theoretical formulation established earlier to extract the corresponding BBN constraints. Using the relation provided in Eq. (\ref{9}), valid within this gravitational setting, we study how the BBN bounds and the Hubble rate $H$ are modified. To study the role of different gravitational contributions, four phenomenological functional forms are considered, each incorporating $R$, $G$, and $\mathcal{T}$ through a different mathematical structure. 
These four models were chosen to explore how changes in geometry and matter can shape the early expansion of the Universe over time. Their equations make it possible to study models of additive curvature and matter terms, direct curvature-matter coupling, and inverse dependence on the Gauss-Bonnet invariant. With these distinct forms, we aim to compare their effects on the freeze-out temperature and primordial helium abundance and to identify the ranges of parameters consistent with BBN observations. These models are used as alternative phenomenological descriptions rather than as unique forms of fundamental gravity theory. Of course. Their physical viability must be judged through the resulting cosmological evolution they produce and whether they are consistent with observational constraints.

\subsection{Model 1}
The first model selected for our examination of BBN constraints is expressed mathematically as
\begin{eqnarray}\label{29}
f(R,G,\mathcal{T})=\alpha_1 R+ G^n+ \gamma_1 \mathcal{T},   
\end{eqnarray}
where $\alpha_1, ~ \gamma_1$ are constants and $n$ is a real exponent that characterizes the nonlinearity of the contribution $G$. This form allows the three sectors to contribute separately to the gravitational dynamics without introducing an explicit coupling between them. The linear dependence on \(R\) retains the standard curvature contribution, while the nonlinear term \(G^n\) provides a higher-curvature modification whose effect can vary with the exponent \(n\). The term proportional to \(\mathcal{T}\) introduces a direct contribution of matter. We therefore use this model as a phenomenological baseline for examining how these three ingredients, acting simultaneously but independently, can modify the expansion rate during the BBN epoch. The reason for choosing this model as our example is to study the effect of the combination of the energy associated with geometry, the non-linear Gauss–Bonnet correction terms, and matter on the expansion rate during the BBN epoch. This phenomenological form can therefore be used to study the impact of the exponent $n$ on the freeze-out temperature and to constrain the parameter space with respect to the observational BBN bounds. It may also be used for a quick check against models with more direct coupling between geometric and matter terms.

To establish the basis for this analysis, the current value of the DE density can be written as
\begin{eqnarray}\label{30}
\Omega_{DE0} = \frac{\rho_{DE0}}{3 M_p^{2} H^{2}}.    
\end{eqnarray}
Here, $\Omega_{DE0}$ represents the current value of the DE density parameter, which is observationally estimated to be around $0.7$, while
\begin{eqnarray}\nonumber
H_0 = 73.02 \pm 1.79 \,\text{km}\,\text{s}^{-1}\,\text{Mpc}^{-1} \;\approx\; 2.1 \times 10^{-42}\,\text{GeV},    
\end{eqnarray}
consistent with the measurements reported in \cite{30,71}.
The optimal estimates for the remaining model parameters are obtained by a combined analysis involving the CC, $H_0$, SNeI and BAO datasets \cite{72}. Assuming that, during the BBN era, the DE density $\rho$ stays effectively constant and equal to its present value $\rho_{DEO}$, we write
\begin{eqnarray}\nonumber
\rho \simeq \rho_{DEO}.  
\end{eqnarray}
This approximation is adopted as a leading-order description under the assumption that the effective DE contribution varies sufficiently slowly over the relevant BBN interval. Since the Universe is radiation-dominated during the BBN epoch, the radiation component provides the dominant contribution to the expansion rate, and the effective DE sector is assumed to remain subdominant. Consequently, any variation of the effective DE density with time is expected to be negligible within the scope of this analysis. However, we emphasize that this assumption is model dependent and is not a general prediction of all $f(R,G,\mathcal{T})$ theories. If the effective DE contribution evolves significantly during the BBN epoch, its time dependence may modify the expansion rate and consequently affect the freeze-out temperature and primordial helium abundance. Hence, the constraints obtained in this work should be interpreted within the approximation of an effective constant or slowly varying DE density during the BBN era.
By substituting Eqs. (\ref{29}), (\ref{6}), and (\ref{7}) into Eq. (\ref{9}), the resulting expression for the energy density $\rho$ can be derived as follows:
\begin{eqnarray}\label{31}
\rho =-\frac{2^{\,1-n} \bigg[2^{4n-1} 3^{\,n} (-H^{4})^{n}-2^{4n-1} 3^{\,2+n} (H^{4})^{n}n+3^{\,n} 4^{\,1+2n} (-H^{4})^{n} n^{2}+ 3 \times 2^{\,n} \alpha_1 H^2\bigg]}{-2 - 3\gamma_1 +\gamma_1\omega}. 
\end{eqnarray}
Putting Eq. (\ref{31}) in Eq. (\ref{30}), we obtain the resulting mathematical expression for the parameter $\gamma_1$.
\begin{eqnarray}\label{32}
\gamma_1=\frac{-2^{3n} 3^{\,n} (-H_0^{4})^{n}+ 2^{3n} 3^{\,2+n} (H_0^{4})^{n} n- 2^{3+3n} 3^{\,n} (-H_0^{4})^{n} n^{2}- 6 H_0^{2} \alpha_1+ 6 H_0^{2} M_p^{2} \Omega_{DEO}}{3 H_0^{2} M_p^{2} (-3 + \omega)\, \Omega_{DEO}}.    
\end{eqnarray}
By inserting Eq. (\ref{32}) into Eq. (\ref{31}) and, upon simplifying Eq. (\ref{27}) with the relation, we derive the final analytical expression.
\begin{eqnarray}\label{33}
\bigg|\frac{\Delta \mathbf{T}_f }{\mathbf{T}_f}\bigg|=\frac{
H_{0}^{2}\left( 6\,\mathbf{T}_f^{4}\,\alpha_1\,\zeta^{2}
+ 24^{n}\,(n-1)\,(8n-1)\left(-\mathbf{T}_f^{8}\,\zeta^{4}\right)^{n} \right)\Omega_{DEO}}{10c_q\,\mathbf{T}_f^{7}\left( 24^{n}\left(-H_{0}^{4}\right)^{n}(n-1)(8n-1)+ 6H_{0}^{2}\alpha_1 \right)\zeta }. 
\end{eqnarray}
Here, The expression for $\mathbf{T}_f$ is taken from Eq. (\ref{21}), whereas the parameter $\zeta$ is specified to be
\begin{eqnarray}\label{34}
 \zeta= \bigg(\frac{4 \pi^3g_*}{45}\bigg)^{1/2} \frac{1}{M_{pl}}.
\end{eqnarray}
\begin{figure}
\centering
\includegraphics[width=0.6\linewidth]{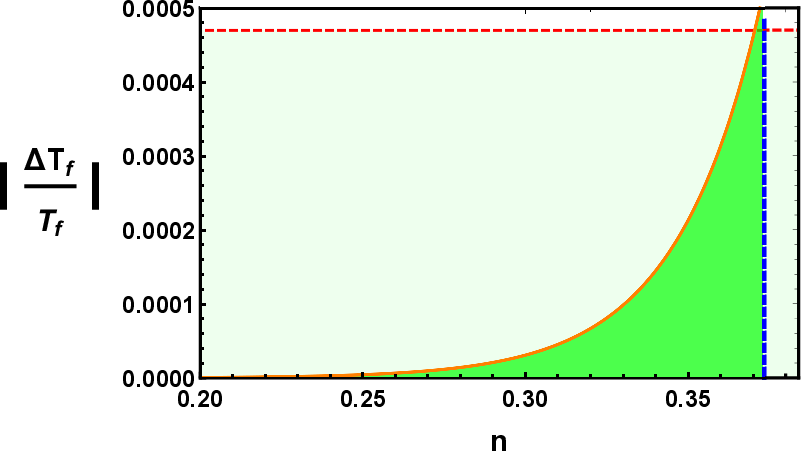}
\caption{Variation of the deviation $\bigg|\frac{\Delta \mathbf{T}_f }{\mathbf{T}_f}\bigg|$ as a function of the free parameter $n$. The shaded region describes the physically admissible range, while the red dashed horizontal line indicates the observational upper bound. The vertical dotted line marks the maximum allowed value of the parameter $n$ compatible with the observational constraints.}\label{m1}
\end{figure}

\textbf{FIG. 1} illustrates how the relative change in the freeze-out temperature, expressed through the ratio $\bigg|\frac{\Delta \mathbf{T}_f }{\mathbf{T}_f}\bigg|$
varies with the exponent $n$ associated with the $G$ term $G^n$ in the model $f(R,G, \mathcal{T})= \alpha_1 R+G^n+\gamma_1 \mathcal{T}$. The parameter $n$ governs the strength of the higher-order curvature corrections introduced by the $G$ contribution, thereby influencing the early Universe expansion rate. The solid blue curve represents the theoretical prediction of $\bigg|\frac{\Delta \mathbf{T}_f }{\mathbf{T}_f}\bigg|$, while the red dashed line denotes the observational upper bound given in Eq. (\ref{28}), inferred from light element abundances, particularly $^4He$ synthesized during BBN. The shaded region below the red dashed line represents the BBN-allowed parameter space, where the predicted freeze-out deviation is consistent with observations. For small values of $n$, the curve lies entirely within this shaded region, indicating negligible deviations from the standard BBN scenario. As $n$ increases, the Gauss–Bonnet contribution becomes increasingly significant, producing rapid growth in $\bigg|\frac{\Delta \mathbf{T}_f }{\mathbf{T}_f}\bigg|$. The curve intersects the observational bound at $n\approx 0.37$. BBN constraints rule out values of $n$ beyond this point (the unshaded region). The analysis is performed using $H_0=70,~ \Omega_{DEO}=0.7$ and $\alpha_1= 10^{-14}$ set to the best fit values. These results demonstrate that BBN imposes a robust upper limit on the $G$ exponent and highlight its effectiveness in constraining extended curvature-based gravity models such as  $f(R,G,\mathcal{T})$. The constraint $n<0.37$ obtained from the present BBN analysis is consistent with observational bounds reported from other cosmological probes, including CMB and large-scale structure studies \cite{73,74}. For the remaining models, the corresponding parameter bounds are slightly larger but remain within observationally acceptable ranges reported in the literature. This indicates that the present BBN analysis provides physically viable, independent constraints on the model parameters during the early Universe epoch.

\subsection{Model 2}
In the second model, we focus on a generalized gravitational action of the form
\begin{eqnarray}\label{35}
f(R,G,\mathcal{T})= \alpha_2 R \,\mathcal{T} +\gamma_2 G^n,  
\end{eqnarray}
where $\alpha_2,~\gamma_2$ are constant coupling parameters and $n$ is a real exponent.
Unlike the additive structure of Model 1, this form is constructed to examine the combined effect of a direct curvature-matter interaction and a nonlinear Gauss-Bonnet contribution. The product $R\mathcal{T}$ allows the matter content to influence the curvature sector explicitly, while the power-law dependence on $G$ provides an independent higher-order geometric modification. This structure is therefore useful for testing whether the interplay between matter-geometry coupling and nonlinear curvature effects can produce an expansion history compatible with the BBN epoch.
The term $\alpha_2 R\,\mathcal{T}$ represents a direct coupling between $R$ and $\mathcal{T}$, making the model sensitive to matter effects during the early radiation-dominated universe. The second part, $\gamma_2 G^n$, incorporates higher-order curvature corrections, where the exponent $n$ determines the degree of deviation from the standard cosmological behavior. Together, these contributions allow the model to capture both matter-geometry coupling and nonlinear geometric effects, providing a framework for testing early Universe dynamics through BBN constraints. The motivation for this model is to study whether an explicit interaction between the curvature and matter along with a nonlinear Gauss-Bonnet contribution can yield an expansion history available from BBN. In particular, this model lets us probe the influence of the coupling parameters and exponent $n$ on the freeze-out temperature and resulting observational constraints. Comparing these results to those of Model 1, we can find the effect of a direct curvature–matter interaction.
By inserting Eqs. (\ref{35}), (\ref{6}), and (\ref{7}) into Eq. (\ref{9}), the corresponding expression for the energy density $\rho$ is obtained in the following form:
\begin{eqnarray}\label{36}
\rho= \frac{2^{ 3n-1}\, 3^{n}\, (-H^{4})^{n}\, \big(1 - 9n + 8n^{2}\big)\, \gamma_2}{\bigg(1 + 3\alpha_2 - 9\alpha_2\, H^2\omega\bigg)} .   
\end{eqnarray}
By inserting Eq. (\ref{36}) in Eq. (\ref{30}), the corresponding closed form expression for the parameter 
$\alpha_2$ is obtained
\begin{eqnarray}\label{37}
\alpha_2=  \frac{-2^{3n} \,3^{n} (-H_0^{4})^{n}\,\gamma_2+ 2^{3n}\, 3^{\,2+n} (-H_0^{4})^{n} n\,\gamma_2 - 2^{3+3n} 3^{n} (-H_0^{4})^{n} n^{2}\,\gamma_2 + 6 H_0^{2}M_p^{2} \Omega_{DEO}}{18 H_0^{4} M_p^{2} ( 3\omega-1)\, \Omega_{DEO}}.  
\end{eqnarray}
By inserting Eq. (\ref{37}) into Eq. (\ref{36}) and then simplifying Eq. (\ref{27}), we arrive at the final analytical expression.
\begin{eqnarray}\nonumber
\bigg|\frac{\Delta \mathbf{T}_f }{\mathbf{T}_f}\bigg|=\frac{
2^{ 3n-1} \, 3^{n} \, H_{0}^{4} \big( n-1\big) \big( 8n-1\big) \gamma_2 \big( -\mathbf{T}_f^{8} \, \zeta^{4} \big)^{n} \, \Omega_{DEO}}{5 c_q \, \mathbf{T}_f^{7} \, \zeta \bigg( 6 H_{0}^{4} M_p^{2} \Omega_{DEO} + \mathbf{T}_f^{4} \zeta^{2} \big( 24^{\, n} (-H_{0}^{4})^{n} ( n-1)(8n-1)\gamma_2 - 6 H_{0}^{2} M_p^{2} \Omega_{DEO}\big)\bigg)}.\\\label{38}  
\end{eqnarray}

\begin{figure}
    \centering
    \includegraphics[width=0.6\linewidth]{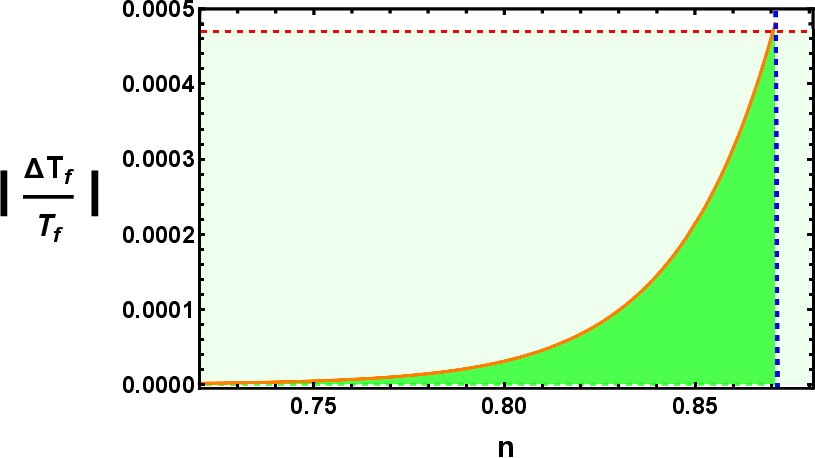}
    \caption{Variation of the ratio $\bigg|\frac{\Delta \mathbf{T}_f }{\mathbf{T}_f}\bigg|$ with respect to the parameter $n$ for Model II.}\label{m2}
\end{figure}
\textbf{FIG. 2} presents the behavior of the relative change in the freeze-out temperature, expressed as $\bigg|\frac{\Delta \mathbf{T}_f }{\mathbf{T}_f}\bigg|$, as a function of the exponent $n$ appearing in the $G$ sector of the second model given in Eq. (\ref{35}). The solid curve represents the theoretical prediction obtained from the modified Friedmann dynamics, while the shaded region highlights the range of $n$ values that satisfy the observational BBN bound on the freeze-out temperature shift. This shaded band therefore denotes the BBN-allowed parameter space, where the predicted light element abundances remain consistent with observations. As $n$ increases, the contribution from the Gauss–Bonnet sector becomes more pronounced, leading to a monotonic growth in $|\Delta \mathbf{T}_f/\mathbf{T}_f|$. Once the curve exits the shaded region, the deviations exceed the observational limit, rendering those values of $n$ incompatible with BBN constraints. In this analysis, the fixed parameters are taken as $ H_0=70,~ \Omega_{DEO}=0.7, ~ M_p=1$ and {\color{blue}$\gamma_2=10^8$} chosen to satisfy the model requirements.
\subsection{Model 3}
In the third model, we adopt a modified gravitational Lagrangian of the form
\begin{eqnarray}\label{39}
f(R,G,\mathcal{T}) = \frac{\alpha_3 R\, \mathcal{T} }{G^n} +\gamma_3, 
\end{eqnarray}
where, $\alpha_3,~ \gamma_3$ are constant parameters. This functional form is constructed to examine the effect of an inverse dependence on the Gauss-Bonnet invariant in the presence of a direct curvature-matter coupling. In contrast to the additive structure of Model 1 and the separate terms in Model 2, the curvature-matter contribution is now modulated by \(G^{-n}\). This structure provides a phenomenological way to investigate whether an inverse Gauss-Bonnet dependence can alter the early Universe expansion while remaining compatible with the stringent requirements of BBN.
This model is fundamentally nonlinear because of the inverse power dependence on the term $G^{-n}$. Even though the term $R\,\mathcal{T}$ is algebraically linear in both curvature and matter, the division by $G^n$ introduces a nonlinear structure in the dynamics. This makes the model significantly different from linear or additive $f(R,G,\mathcal{T})$ forms and allows us to explore how inverse $G$ corrections modify the early-Universe expansion. The constant term $\gamma_3$ acts as an effective cosmological contribution and does not affect the nonlinearity of the model. The idea to study this model is due to the hopes to discover how the BBN results of curvature-matter coupling changes when we have an inverse dependence on the Gauss-Bonnet invariant. Unlike before, the strength of the $R\mathcal{T}$ contribution is now explicitly dependent on $G^{-n}$, which lets us study the effects of this dependence on the cosmic expansion rate and freeze-out temperature. Constraints derived from this procedure foresee ably determine whether the model's parameters are consistent with observational BBN limits. By substituting Eqs. (\ref{39}), (\ref{6}), and (\ref{7}) into Eq. (\ref{9}), the resulting form of the energy density $\rho$ can be expressed as follows
\begin{eqnarray}\label{40}
\rho=  \frac{2^{ 4n-1} \, 3^{n} \left(-H^{4}\right)^{n}  \gamma_3}{48^{n} \left(-H^{4}\right)^{n}+ H^2 \bigg( 3 \times 2^{\,n} \alpha_3 + 3 \times 2^{\,3+n} n \alpha_3- 9 \times 2^{\,n} \alpha_3 \omega- 9 \times 2^{\,3+n} n \,\alpha_3\, \omega\bigg)} . 
\end{eqnarray}
Using Eq. (\ref{40}) within Eq. (\ref{30}), the resulting closed form expression for the parameter $\alpha_3$ is obtained
\begin{eqnarray}\label{41}
\alpha_3 = \frac{2^{3n-1} \times 3^{ n-2} (H_0^{4})^{n}
\left( -\gamma_3 + 6 H_0^{2} M_p^{2} \Omega_{DEO}\right)}{H_0^4(1+8n) M_p^2(3\omega-1)\Omega_{DEO}}.
\end{eqnarray}
Upon inserting Eq. (\ref{41}) into Eq. (\ref{40}) and continue the simplification of Eq. (\ref{27}), the final analytical expression is derived
\begin{eqnarray}\label{42}
\bigg|\frac{\Delta \mathbf{T}_f }{\mathbf{T}_f}\bigg|= \frac{\gamma_3}{10\, c_q\, \mathbf{T}_f^{7}\, \zeta \left(
6M_p^{2}+\frac{\left(-H_{0}^{4}\right)^{n}\, \mathbf{T}_f^{4}\, \zeta^{2}\,\left(-\mathbf{T}_f^{8}\, \zeta^{4}\right)^{-n}\left(\gamma_3 - 6 H_{0}^{2} M_p^{2} \Omega_{DEO} \right)}{H_{0}^{4}\, \Omega_{DEO}}\right)}. 
\end{eqnarray}
\begin{figure}
    \centering
    \includegraphics[width=0.6\linewidth]{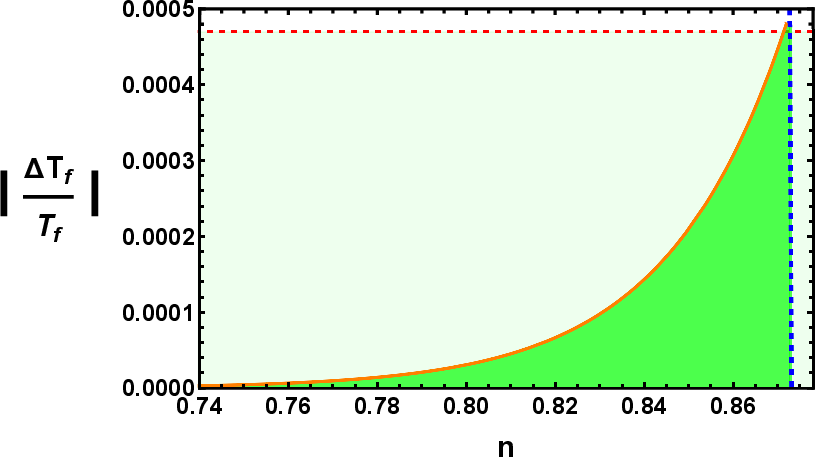}
    \caption {Behavior of the ratio $\bigg|\frac{\Delta \mathbf{T}_f }{\mathbf{T}_f}\bigg|$ as a function of the parameter $n$ in Model III.}
\end{figure}
\textbf{FIG. 3} demonstrate the behavior of $\bigg|\frac{\Delta \mathbf{T}_f }{\mathbf{T}_f}\bigg|$ as a function of the free parameter $n$, based on the analytical relation obtained in Eq. (\ref{42}). For this analysis, the fixed parameters are chosen as $H_0=70,~ \Omega_{DEO}=0.7,~ M_p=1$ and $\gamma_3 = 10^{12}$, while the parameter 
$n$ is continuously varied within the allowed range. The solid curve represents the theoretical prediction of the model, while the shaded region highlights the range allowed by BBN observational constraints. For smaller values of $n$, the predicted deviation remains well within the shaded region, indicating consistency with primordial light element abundances. As $n$ increases, $|\Delta \mathbf{T}_f/\mathbf{T}_f|$ rises rapidly and eventually exits the shaded band, signaling a violation of BBN bounds.
\subsection{Model 4}
In the fourth model of our modified gravity framework, we consider a function defined by
\begin{eqnarray}\label{43}
f(R,G,\mathcal{T})= \alpha_4 R +\frac{\beta_1 \mathcal{T}}{G^n}+\gamma_4, 
\end{eqnarray}
where, $\alpha_4,~\beta_1$ and $\gamma_4$ are constant parameters. This functional form is constructed to examine an alternative way in which the matter and Gauss-Bonnet sectors can contribute to the gravitational dynamics. In contrast to the previous models, the matter contribution is placed over a power of the Gauss-Bonnet invariant, so that its effect is directly modulated by the geometric quantity \(G\). This structure provides a phenomenological test of whether an inverse Gauss-Bonnet dependence in the matter sector can produce a viable modification of the early Universe expansion while remaining consistent with BBN constraints. This model combines both linear and nonlinear components. The first term, $\alpha_4 R$, represents a linear dependence on $R$, similar to the standard GR contribution but scaled by a constant coupling. The second term introduces a nonlinear dependence involving $\mathcal{T}$ and $G$ through an inverse power of $G$, making the framework sensitive to higher-order curvature effects. The additive constant $\gamma_4$ serves as a background geometrical contribution, effectively acting as a generalized cosmological constant. Due to the inverse $G$ dependence $G^{-n}$, this model exhibits nontrivial nonlinear dynamics, allowing us to examine potential deviations from GR during the early Universe, particularly in the regime relevant to BBN. The motivation for this choice of model is to address the question of the role of a geometry-dependent matter contribution as it pertains to the expansion history in the early Universe. In this model, an inverse power of the Gauss–Bonnet invariant is added to the matter sector to provide yet another means of studying departures from standard cosmology. Our main attention here is to study the influence of the model parameters on the freeze-out conditions and verify whether the emerging predictions are aligned with the constraints inferred observationally at the BBN era.
Upon substituting Eqs. (\ref{43}), (\ref{6}), and (\ref{7}) into Eq. (\ref{9}), the energy density $\rho$ reduces to the following analytic form:
\begin{eqnarray}\label{44}
\rho= \frac{2^{\,1+4n}\, 3^{n}\, H^{2}\, \left(-H^{4}\right)^{n}
\left( 3\alpha_4 + \frac{\gamma_4}{2H^{2}} \right)
}{2^{\,1+4n}\, 3^{n}\, (-H^{4})^{n}+ 9\times 2^{n} n \beta_1+ 2^{\,3+n} n^{2} \beta_1
- 2^{n} \beta_1 \omega(1+ 27 n)+ 3 \beta_1(2^{n}-  2^{\,3+n} n^{2} \omega)}.   
\end{eqnarray}
By applying Eq. (\ref{44}) to Eq. (\ref{30}), we obtain the explicit closed-form expression for the parameter $\beta_1$
\begin{eqnarray}\label{45}
\beta_1= \frac{2^{3n}\times 3^{\,n-1}\, (-H_0^{4})^{\,n}\,\left(-6H_0^{2}\alpha_4 - \gamma_4 + 6H_0^{2} M_p^{2}\Omega_{DEO}\right)}{H_0^{2} M_p^{2}\,\big( \omega + 27n\,\omega + 24n^{2}\omega-3-9n-8 n^2\big)\,\Omega_{DEO}}. 
\end{eqnarray}
Using Eq. (\ref{45}) within Eq. (\ref{44}) and simplifying Eq. (\ref{27}) thereafter, we derive the resulting closed-form analytical expression
\begin{eqnarray}\label{46}
\bigg|\frac{\Delta \mathbf{T}_f }{\mathbf{T}_f}\bigg|=\frac{
H_0^{2}\,(-\mathbf{T}_f^{8}\,\zeta^{4})^{n}\,(\gamma_4 + 6\,\mathbf{T}_f^{4}\alpha_4\,\zeta^{2})\,\Omega_{DEO}
}{10\,c_q\,\mathbf{T}_f^{7}\,\zeta \left(6H_0^{2}M_p^{2}(-\mathbf{T}_f^{8}\zeta^{4})^{n}\Omega_{DEO}+ (H_0^{4})^{n}\left(\gamma_4 + 6H_0^{2}(\alpha_4 - M_p^{2}\Omega_{DEO})\right)\right)}.   
\end{eqnarray}
\begin{figure}
    \centering
    \includegraphics[width=0.6\linewidth]{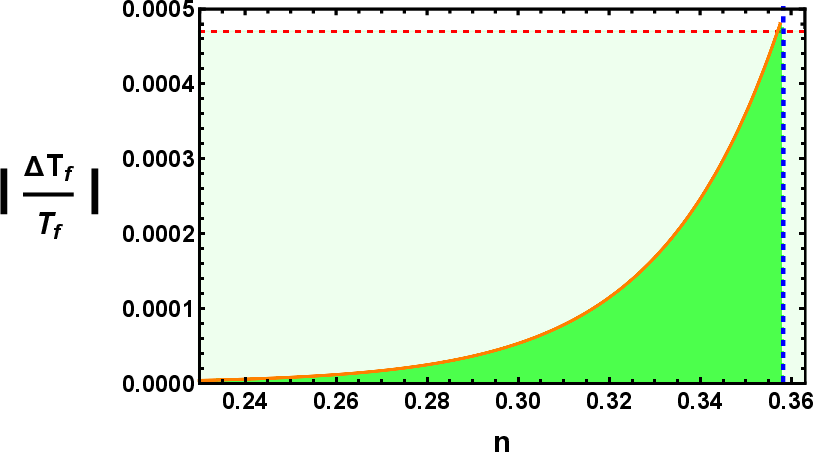}
    \caption{Variation of the ratio $\bigg|\frac{\Delta \mathbf{T}_f }{\mathbf{T}_f}\bigg|$ against the parameter $n$ within Model IV.}
\end{figure}
\textbf{FIG. 4} presents the variation of the freeze-out temperature ratio $\bigg|\frac{\Delta \mathbf{T}_f }{\mathbf{T}_f}\bigg|$ with respect to the model parameter 
$n$. For the numerical analysis of this model, we set $H_0=70,~ \Omega_{DEO}=0.7,~M_p=1$. The model parameters are fixed to $\alpha_4=0.1$ and $\gamma_4= 10^{12}$. The solid curve represents the theoretical prediction, while the horizontal dashed line corresponds to the observational upper bound imposed by BBN. The shaded region highlights the range of values $n$ for which $|\Delta \mathbf{T}_f/\mathbf{T}_f|$ remains below the observational limit, indicating consistency with the light element abundance data. As 
$n$ increases, the deviation grows rapidly, and values beyond the shaded region are excluded since they lead to freeze-out temperature shifts incompatible with BBN constraints.

\section{Primordial Helium Abundance Constraints }

In this section, we extend our BBN analysis by examining the primordial $^4He$ mass fraction $Y_p$ given in Eq. (\ref{22}) within the framework of $f(R,G,\mathcal{T})$ gravity. The abundance of $^4$He is one of the most sensitive probes of the early Universe expansion rate, as it is directly governed by the neutron–proton freeze-out process and the subsequent nuclear reaction history. Any deviation from the standard expansion dynamics can leave a measurable imprint on $Y_p$. Building on the four $f(R,G,\mathcal{T})$ gravity models introduced in the previous section, we examine how the modified Friedmann equations affect $He$ production during the radiation-dominated era. For each model, the theoretical predictions of $Y_p$ are compared with the observational bound given in Eq. (\ref{23}).

\begin{figure}
    \centering
    \includegraphics[width=0.6\linewidth]{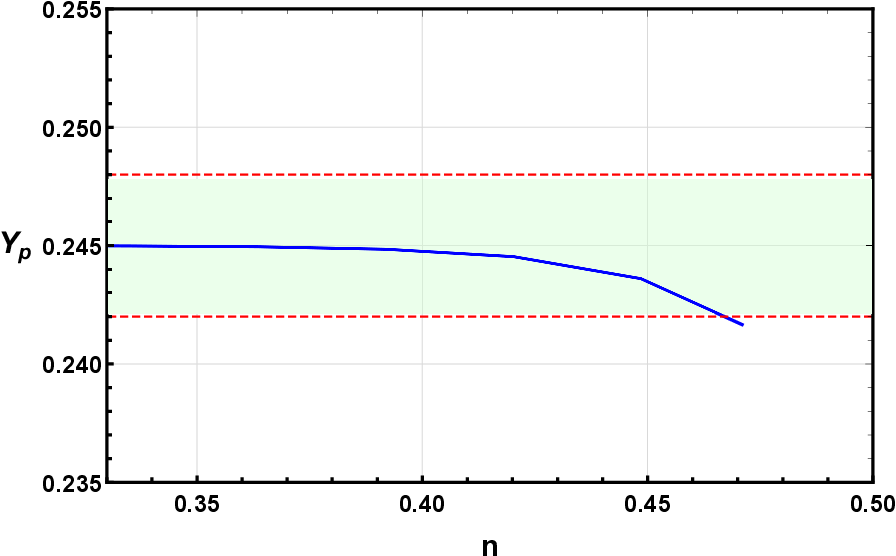}
    \caption{Variation of the primordial $He$ mass fraction $Y_p$ with respect to the model parameter $n$ for model 1.}
    \label{fig:placeholder}
\end{figure}
\textbf{FIG. 5} displays the variation of the primordial $He$ mass fraction $Y_p$ with respect to the model parameter $n$ for the first model given in Eq. (\ref{29}). The solid blue curve represents the theoretical prediction obtained within the gravity framework $f(R,G,\mathcal{T})$, while the light-shaded area denotes the observational limit inferred from measurements of the primordial element.
For smaller parameter $n$ values, the predicted $Y_p$ remains well inside the shaded region for a limited parameter range, confirming consistency with BBN. Outside this range, the predicted $He$ abundance exceeds the observational limit. The analysis is performed at $\alpha_1 = 10^{-14}$, where the theoretical prediction for $Y_p$ remains within the BBN observational limits.

\begin{figure}
    \centering
    \includegraphics[width=0.6\linewidth]{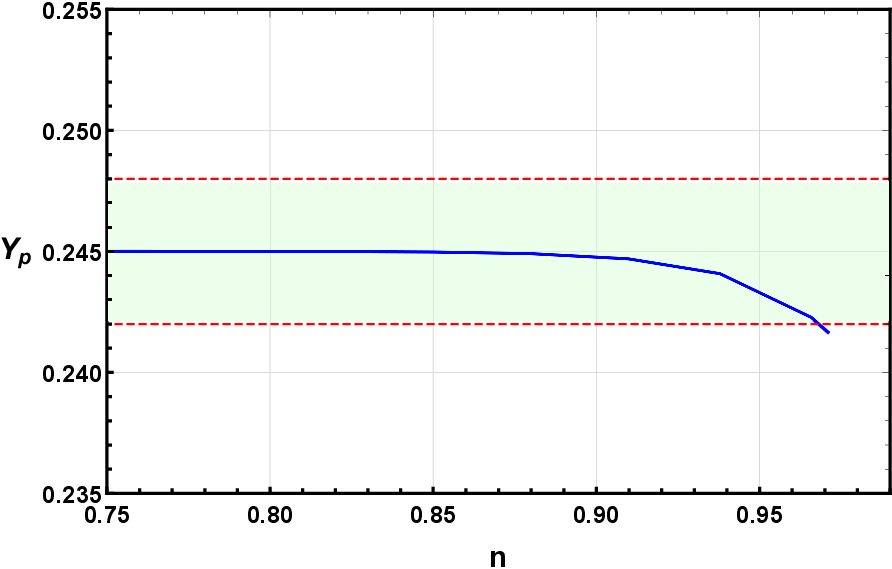}
    \caption{Dependence of the primordial $He$ mass fraction $Y_p$ on the parameter $n$ for the model 2 within the $f(R,G,\mathcal{T})$ gravity framework.}
    \label{fig:placeholder}
\end{figure}
 
\textbf{FIG. 6}  illustrates the dependence of the $^4He$ abundance $Y_p$ on the free model parameter $n$ for the second model given in Eq. (\ref{35}). The shaded region denotes the observationally allowed range inferred from primordial $He$ abundance measurements. The solid blue curve represents the theoretical prediction of the model and shows a smooth and monotonic variation with the model parameter. For smaller values of the parameter, the blue curve remains entirely confined within the shaded band, indicating full consistency with BBN constraints. As the parameter increases further, the curve gradually approaches and eventually crosses the observational boundary. This behavior demonstrates the strong sensitivity of $He$ production to the model parameters. The analysis is performed by fixing the constant at $\gamma_2 = 10^8$.

\begin{figure}
    \centering
    \includegraphics[width=0.6\linewidth]{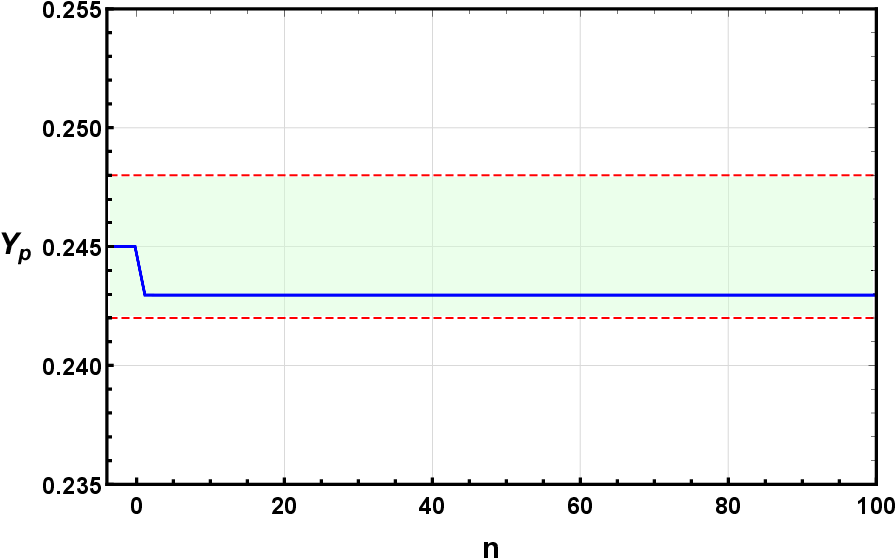}
    \caption{Evolution of the $He$ abundance $Y_p$ as a function of the model parameter $n$ for the third $f(R,G,\mathcal{T})$ gravity model.}
\end{figure}

\textbf{FIG. 7} shows the behavior of the $He$ mass fraction $Y_p$ as a function of the parameter $n$ for model 3 given in Eq. (\ref{39}) at $\gamma_3=10^12$. The blue curve first decreases slightly and then increases with $n$, indicating a transition in the impact of the nonlinear Gauss–Bonnet–matter coupling on $He$ production. The shaded region denotes the observationally allowed range of 
$Y_p$, and the model remains consistent with the BBN constraints as long as the curve lies within this band.

\begin{figure}
    \centering
    \includegraphics[width=0.6\linewidth]{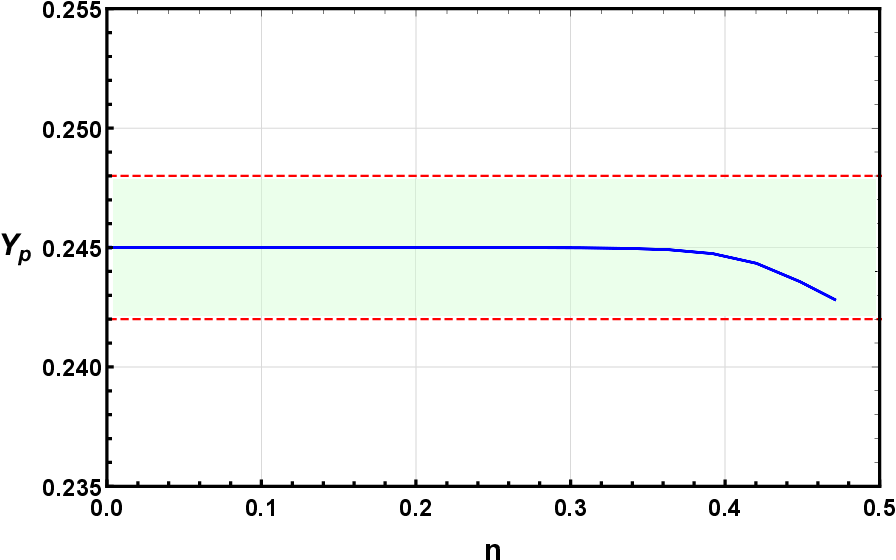}
    \caption{$He$ mass fraction $Y_p$ as a function of the parameter $n$ for the fourth gravity model.}
    \label{fig:placeholder}
\end{figure}
 \textbf{FIG. 8} shows the variation of the $He$ mass fraction 
$Y_p$ with the model parameter $n$ for model 4 given in Eq. (\ref{43}), evaluated at $\gamma_4=10^{12}$ and $\alpha_4=0.1$. The solid blue curve represents the theoretical prediction, while the shaded band denotes the observationally allowed region from BBN. For smaller values of $n$, the predicted $Y_p$ remains well within the observational bounds, indicating consistency with BBN. However, as the parameter $n$ is increased, the blue curve crosses the shaded region, signaling a violation of BBN constraints. This behavior demonstrates the strong sensitivity of $He$ abundance to the coupling constant $\alpha$ in this model.

\begin{table*}[ht]
\centering 
\caption{Summary of the parameter ranges allowed by the freeze-out temperature and primordial helium abundance constraints, together with the resulting common compatible regions for the four $f(R,G,\mathcal{T})$ models.} 
\label{5a}

\begin{tabular}{p{5.3cm} p{5.3cm}  p{5.3cm} }
  \hline
  \hline

 ~~~~ \textbf{$\frac{\Delta T_f}{T_f}$ Constraints} &  ~~~~\textbf{$Y_p$ Constraints} & ~~~~\textbf{Common Region} \\
  
  \hline
  \hline
  \multicolumn{3}{c}{\textbf{Model 1}} \\
  
  \hline
  \hline
  
  ~~~~$-10^{-3}<\alpha_1<10^{-3}$ &~~~$-10^{-12}<\alpha_1<10^{-12}$&  ~~~$-10^{-12}<\alpha_1<10^{-12}$ \\
   ~~~~~~$0<n<0.37$ &   ~~~~~~$0.32<n<0.47$&   ~~~~~~$0.32<n<0.37$ \\
  
  \hline
  \hline
  \multicolumn{3}{c}{\textbf{Model 2}} \\
  
  \hline
  \hline
  
  ~~~$-10^{302}<\gamma_2<10^{302}$ &   ~~~~~$10^{6}<\gamma_2<10^{298}$&  ~~~~~$10^{6}<\gamma_2<10^{298}$ \\
   ~~~~~~$0<n<0.872$ &   ~~~~~$0.529<n<0.968$&   ~~~~~$0.529<n<0.872$ \\
  
  \hline
  \hline

  \multicolumn{3}{c}{\textbf{Model 3}} \\
  
  \hline
  \hline
  
  ~~~~~$0<\gamma_3<10^{331}$ &   ~~~~~$0<\gamma_3<1.2 \times 10^{12}$&   ~~~~~$0<\gamma_3<1.2 \times 10^{12}$ \\
    ~~~~~~$0<n<0.872$ &   ~~~~~$n>0$&   ~~~~~$0<n<0.872$ \\
  
  \hline
  \hline

  \multicolumn{3}{c}{\textbf{Model 4}} \\
  
  \hline
  \hline
  
  ~~~$-\infty<\alpha_4<\infty$ &   ~~~~~$0<\alpha_4<0.21$&   ~~~~~$0<\alpha_4<0.21$ \\
  ~~~~~~$0<\gamma_4<10^{14}$ &   ~~~~~$10^6<\gamma_4<10^{14}$&   ~~~~~$10^{6}<\gamma_4<10^{14}$ \\
   ~~~~~~$n<0.357$ &   ~~~~~$0<n<0.471$&   ~~~~~$0<n<0.357$ \\
  
  \hline
  \hline
  
\end{tabular}
\end{table*}

A comparative evaluation of the allowed parameter ranges reveals varying degrees of parameter constraints among the four models. The model {\bf1} is the most tightly
constrained, with $\alpha_1$ restricted to the narrow interval
$-10^{-3}<\alpha_1<10^{-3}$ and $n$ limited to $0.32<n<0.37$ in the
common region, indicating a comparatively stronger degree of parameter
fine-tuning. In contrast, Model {\bf2} allows for a relatively wide range of parameters
$\gamma_2$, namely $10^{6}<\gamma_2<10^{298}$, and can therefore be regarded
as the least constrained with respect to this parameter. The models {\bf3} and {\bf4}
also allow for relatively broad feasible regions. In particular, Model 4 admits
$0<\alpha_4<0.21$ and $10^{6}<\gamma_4<10^{14}$ in the common region,
together with $n<0.357$. Overall, the results indicate that the four
models allow different ranges of parameter restriction,
with Model 1 requiring the strongest tuning, while Models {\bf2-4} retain
broader regions compatible with the BBN constraints.

\section{Concluding Remarks}

In this study, we examined the implications of $f(R,G,\mathcal{T})$ gravity for the physics of BBN, which is extremely sensitive to the expansion rate of the early Universe. The main novelty of this work is the systematic investigation of BBN constraints for four phenomenological functional forms $f(R,G,\mathcal{T})$, constructed to explore different combinations of curvature, Gauss-Bonnet, and matter-geometry contributions. In particular, both the freeze-out temperature shift and the primordial helium abundance $Y_p$ constrain the same model parameter space, providing complementary tests of these models' viability during the BBN epoch. This analysis identifies the parameter regions in which the considered modified-gravity scenarios remain compatible with primordial nucleosynthesis observations. It provides a precise way to test whether any modified gravity theory can remain consistent with the observed abundances of light elements. To evaluate the viability of the theory, we calculated the relative change in the freeze-out temperature for different parameter choices and compared these results with the observational limit $\bigg|\frac{\Delta \mathbf{T}_f }{\mathbf{T}_f}\bigg|<4.7 \times10^{-4}$. In addition to the freeze-out analysis, we also investigated the primordial $He$ mass fraction $Y_p$ for all four models, providing an independent and complementary constraint on the theory. The abundance results of $He$ show that the predicted values remain within the observed range only for restricted parameter intervals, further tightening the limits on the model parameters. Our analysis covers four separate models of the  $f(R,G,\mathcal{T})$ framework. For all four models, we adopt the same background parameters, choosing $H_0=70,~ \Omega_{DEO}=0.7$
and throughout the analysis.

\textbf{Model 1} introduces a combined contribution of the $R$, $G$ and $\mathcal{T}$ through additive terms controlled by the constants $\alpha_1 \gamma_1$ and exponent $n$. The interaction among these curvature and matter terms modifies the expansion rate during the BBN epoch. By comparing the resulting freeze-out temperature shift with observational bounds, we find that the model remains consistent with BBN only when the parameter $n$ lies in the interval $(0,0.37)$. This allowed interval is highlighted by the shaded region in the figure, where the theoretical curve remains below the observational bound. Within this domain, the freeze-out temperature deviations are sufficiently small, ensuring that the resulting light element abundances remain consistent with BBN observations.
In \textbf{Model 2}, a nonlinear multiplicative term $\alpha R\,\mathcal{T}$ is combined with an additional $G$ contribution $\gamma_2 G^n$, allowing both curvature matter coupling and higher-order geometry to modify the expansion rate. The shaded area clearly illustrates the model's narrow viability range and shows how sensitively early-Universe thermodynamics constrains the $G$ coupling through primordial nucleosynthesis, as shown in \textbf{FIG. 2}. For numerical analysis, we adopt the constant $\gamma_2 = 10^8$. 
\textbf{Model 3} introduces a nonlinear structure through the multiplicative term $\alpha_3 R \, \mathcal{T}$ divided by the $G^n$, along with a constant shift. The inverse power of $G$ amplifies the model’s sensitivity to early-Universe dynamics, causing the freeze-out temperature to depend strongly on $n$. \textbf{FIG. 3} clearly demonstrates that only a narrow interval of $n$ is permitted, emphasizing the strong constraining power of BBN on the model parameters. For this analysis, the constant parameter is fixed at $\gamma_3= 10^{12}$. 
\textbf{Model 4} introduces a nonlinear structure through the additive combination of the $R$ and the curvature term of matter $\beta_1\, \mathcal{T}/G^n$, together with the constant $\gamma_4$. The inverse dependence $G^{-n}$ greatly increases the model’s sensitivity to early Universe expansion, making the freeze-out temperature strongly dependent on $n$. From comparison with the BBN bounds, the model is found to be viable only within the shaded interval of $n$ in \textbf{FIG. 4}, where the theoretical curve remains under observational constraint. This region represents the parameter space in which the predicted freeze-out temperature and light element yields remain compatible with BBN data. We adopt constant parameters $\alpha_4=0.1$ and $\gamma_4= 10^{12}$.

Our analysis demonstrates that all selected models remain compatible with observationally allowed ranges of the freeze-out temperature, provided the parameter $n$ is restricted to a narrow, physically meaningful interval, as illustrated in \textbf{FIG. 1,2,3} and \textbf{4}. Within these bounds, the modified cosmological dynamics introduced by the additional curvature matter couplings do not disturb the thermal history required for successful light element formation. Overall, the $f(R,G,\mathcal{T})$ framework shows consistency with the BBN limits and therefore represents a viable extension of standard cosmology at the beginning of the period.

We conclude that all four considered models within the $f(R,G,\mathcal{T})$ gravity framework are consistent with the observational constraints on the primordial $He$ mass fraction 
$Y_p$ given in Eq. (\ref{23}). For each model, the theoretical predictions of $Y_p$ remain within the observationally allowed shaded region for suitable ranges of the free parameter $n$ as shown in \textbf{FIG. 5,6,7} and \textbf{8}. The graphical analysis indicates that the predicted $He$ abundance remains consistent with observations for specific allowed parameter values. However, as the model parameters are varied beyond these permitted ranges, the theoretical curves cross the observationally allowed band, leading to noticeable deviations from the standard primordial $He$ fraction. This demonstrates that, despite the presence of higher-order curvature and matter–geometry coupling terms, the theory can successfully reproduce the observed helium-4 abundance produced during BBN. Therefore, the $Y_p$ analysis provides strong complementary evidence, alongside the freeze-out temperature constraints, that $f(R,G,\mathcal{T})$ gravity remains a viable extension of GR at early cosmological times.

The parameter ranges obtained in this work are constrained from the perspective of primordial nucleosynthesis; however, their complete cosmological viability should also be examined through observations associated with other stages of the Universe. The BBN-compatible regions identified here provide an important early Universe restriction on the considered $f(R,G,\mathcal{T})$ models, while additional observational probes may further test these ranges. For example, CMB measurements can provide information about the cosmological evolution around the recombination era, whereas BAO and CC observations can place further restrictions on the late-time expansion history. In addition, the modifications introduced by the gravitational sector may affect the evolution of matter perturbations and the growth of large-scale structures. Therefore, an analysis of cosmological perturbations and structure formation would be useful for determining whether the parameter regions allowed by BBN remain viable under a broader set of cosmological observations. Such investigations are beyond the scope of the present work but represent a natural extension of the current analysis.

Beyond the scope of the present investigation, several promising directions remain open for further study. The models considered here may be examined using additional early Universe probes, including CMB constraints, the baryon-to-photon ratio, and primordial gravitational waves. Combining these complementary observations with the BBN bounds obtained in the present work could provide a broader assessment of the cosmological viability of $f(R,G,\mathcal{T})$  gravity and its potential role in describing the evolution of the universe across different cosmological epochs.

\section*{Acknowledgements}
Princess Nourah bint Abdulrahman University Researchers Supporting Project number (PNURSP2026R59), Princess Nourah bint Abdulrahman University, Riyadh, Saudi Arabia.

\section*{Data Availability Statement}
This study is purely theoretical and does not involve any associated datasets. Hence, data sharing is not applicable.

\section*{Conflict of Interest}
Authors declare that they have no conflict of interest.

\vspace{.25cm}

\begin{thebibliography}{36}



\bibitem{2} S. Perlmutter et al.; Measurements of and from 42 high-redshift supernovae. $Astrophy.~ J.$ \textbf{517}, 565 (1999).

\bibitem{3} D. N. Spergel et al.; Wilkinson microwave anisotropy probe (WMAP) three year results: implications for cosmology. $Astrophys~ J. ~Suppl.$ \textbf{170}, 377 (2007). 

\bibitem{4} W. Hu, S. Dodelson.; Cosmic microwave background anisotropies. $Annu. ~Rev.~ Astron.~ Astrophys.$ \textbf{40}, 171 (2002).


\bibitem{6} S. Nojiri and S. D. Odintsov.; Introduction to modified gravity and gravitational alternative for dark energy. $Int. ~J.~ Geom.~ Meth.~ Mod.~ Phys.$ \textbf{4}, 115 (2007).

\bibitem{7} K. A. Olive, G. Steigman T. P. Walker.; Primordial nucleosynthesis: Theory and observations. $Phys.~Rep$ \textbf{333}, 389 (2000). 


\bibitem{9} R. H. Cyburt, B. D. Fields,  K. A. Olive and T. H. Yeh.; Big bang nucleosynthesis: Present status. $Rev.~Mod.~Phys.$ \textbf{88}, 015004 (2016).

\bibitem{10} A. Einstein.; Sitzungsberichte der Preussischen Akademieder Wissenschaften zu Berlin. $Die~Feldgleichungun~der~ Gravitation$ \textbf{25}, 844 (1915).

\bibitem{11} D. Hilbert.; Nachrichten von der Gesellschaft der Wissenschaften zu Göttingen Mathematisch - Physikalische Klasse 3. $Die ~Grundlagen ~der ~Physik$ \textbf{1915}, 395  (1915).

\bibitem{12} R. A. Alpher, H. Bethe, G. Gamow.; The origin of chemical elements. $Phys.~Rev.$ \textbf{73}, 803 (1948).

\bibitem{13} E. W. Kolb and M. S. Turner, The Early Universe, $Addison~Wesley$ (1990).

\bibitem{14} A. De Felice, S. Tsujikawa.; $f(R)$ Theories. $Living ~Rev.~ Rel.$ \textbf{13}, 3 (2010).

\bibitem{15} R. Ferraro, F. Fiorini.; Modified teleparallel gravity: Inflation without inflaton. $Phys.~ Rev. ~D$ \textbf{75}, 084031 (2007).

\bibitem{17} Y. Fujii, K. Maeda.; The Scalar Tensor Theory of Gravitation. $Cambridge~ University~ Press$, ISBN:0-511-02988-8 (2003).

\bibitem{18} J. Lu, X. X. Zhao, G. Chee.; Cosmology in symmetric teleparallel gravity and its $f(Q)$ extension. $Eur.~ Phys.~ J.~ C$ \textbf{79}, 530 (2019).

\bibitem{21} N. Myrzakulov,  S. H. Shekh and A. Pradhan.; Cosmological implications of $f(R,\Sigma,\mathcal{T})$ gravity: A unified approach using OHD and SNIa data. $Phys.~ Lett.~ B$ \textbf{862} 139369 (2025).

\bibitem{22} L. V. Jaybhaye, R. Solanki, S. Mandal, P. K. Sahoo.; Cosmology in $f(R, L_m)$ gravity. $Phys.~ Lett.~ B$ \textbf{831}, 137148 (2022).

\bibitem{25} T. B. Gonc¸alves, L. Atayde  and N. Frusciante.; Cosmological study of a symmetric teleparallel gravity model. $Phys.~ Rev.~ D$ \textbf{109(8)}, 084003 (2024).

\bibitem{26} J. K. Singh, H. Balhara, K. Bamba, J. Jena.; Cosmic analysis of a model in higher-order gravity theory. $Astro.~ Computing$ \textbf{46}, 100790 (2024).

\bibitem{28} Q. Wang, X. Ren, Y.F. Cai, W. Luo and E. N Saridakis.; Observational Test of $f(Q)$ Gravity with Weak Gravitational Lensing. $The~ Astrophysical ~Journal$ \textbf{974}, 7 (2024).

\bibitem{29} S. Capozziello, G. Lambiase and E. N. Saridakis.; Constraining $f(T)$ teleparallel gravity by big bang nucleosynthesis. $Eur.~ Phys. ~J. ~C$ \textbf{77}, 576 (2017).


\bibitem{30}  S. Bhattacharjee, P. K. Sahoo.; Big bang nucleosynthesis and entropy evolution in $f(R,T)$ gravity. $Eur.~ Phys.~ J.~ Plus$ \textbf{135}, 350 (2020).

\bibitem{31} J. D. Barrow, S. Basilakos, E. N. Saridakis.; Big bang nucleosynthesis constraints on Barrow entropy. $Phys.~ Lett.~ B$ \textbf{815}, 136134 (2021).

\bibitem{32}  P. Asimakis, et al.; Big bang nucleosynthesis constraints on $f(T,T_G)$ gravity. $Universe$ \textbf{8}, 486 (2022).

\bibitem{33} P. Asimakis, et al.; Big bang nucleosynthesis is constraints on higher order modified gravities. $Phys. ~Rev.~ D$ \textbf{105}, 084010 (2022).

\bibitem{34} S. Bhattacharjee.; BBN constraints on $f(Q, T)$ gravity. $Int. ~J.~ Mod.~ Phys.~ A$ \textbf{37}, 2250017 (2022).


\bibitem{35} F. K. Anagnostopoulos, V.  Gakis, E. N. Saridakis, S. Basilakos.; New models and big bang nucleosynthesis constraints in $f(Q)$ gravity. $Eur.~ Phys.~ J.~ C$ \textbf{83}, 58 (2023).

\bibitem{36}  A. Giri, R. J. Scherrer.; Big bang nucleosynthesis with rapidly varying $G$. $Phys. ~Rev.~ D$ \textbf{109}, 103521 (2024).

\bibitem{37} J. Ge, L. Ming, S. D. Liang, H. H. Zhang, T. Harko.; Constraining Weyl type $f(Q,T)$ gravity with Big Bang Nucleosynthesis. $Phy.~ Rev.~ D$ \textbf{111}, 124049 (2025).

\bibitem{38} A. Boccia, F. Iocco, L. Visinelli.;  Constraining the primordial black hole abundance through Big-Bang nucleosynthesis. $Phy.~ Rev.~ D$ \textbf{111}, 063508 (2025).

\bibitem{39} B. Laminel, et al.; Cosmological measurement of the gravitational constant $G$ using the CMB, BAO, and BBN. $A ~\& ~A$ \textbf{697}, A109 (2025).

\bibitem{40} A. M. Sultan, M. Ali, S. Rani, N. Azhar, N. Myrzakulovf and S. Shaymatovg.;  Constraining Big Bang nucleosynthesis in $f(T,B,T_\mathcal{G},B_\mathcal{G})$ gravity. $Nucl. ~Phys.~ B$ \textbf{1018}, 117023 (2025).

\bibitem{41} D. Jang, M. R. Gangopadhyay, M-Ki Cheoun,  T. Kajino, M. Sam.; Big Bang Nucleosynthesis constraints on the Energy-Momentum Squared Gravity: The $T^2$ model. $Phy.~ Rev. ~D$ \textbf{111},  043525 (2025).

\bibitem{42} A. M. Sultan, M. Fatima, J. L. Said.; BBN in Constrained $f(T, \varphi)$ Gravity Through Various Observational Schemes. Class. $Quan.~ Grav.$ \textbf{42}, 175009 (2025).

\bibitem{43} Y. Al-Omar, M. Nahili.; Synergistic constraints on extensions of telleparallel gravity from primordial nucleosynthesis and cosmic chronometers. $Phys.~ Scr.$ \textbf{100}, 095010 (2025).

\bibitem{44} T. M. Matei, C. A. Croitoru, T. Harko.; Big Bang Nucleosynthesis constraints on the cosmological evolution in a Universe with a Weylian boundary. $Eur. ~Phys.~ J.~ C.$ \textbf{85}, 1092 (2025).

\bibitem{45} A. Sheikh, A. Shabahzi.; Barrow Cosmology and Big-Bang Nucleosynthesis. $Phys. ~Rev.~ D$ \textbf{111},  043518, (2025).

\bibitem{46} P. Braat and M. Hufnagel.; Big Bang Nucleosynthesis constraints on resonant DM
annihilations. $JCAP$ \textbf{02}, 032 (2025).

\bibitem{47} S. S. Luo, Q. Q. Jiang, Z. W. Feng, X. Zhou and X. L.  Mu.;  Effects of a Higher-Order Generalized Uncertainty Principle on Big Bang Nucleosynthesis. $Eur.~ Phys.$ \textbf{140}, 331 (2025).

\bibitem{47a} A. M. Sultan, M. Fatima, J. L. Said, A. Batool.; Constraining big bang nucleosynthesis in $f (T, B)$ gravity through observational analysis. $Phys.~ Dark~ Univ.$ \textbf{49},  102023 (2025).

\bibitem{47b} A. Eid, M. A. Ibrahem, A. M. Sultan, M. Ali, M. U. Shahzad, H. U. Rehman.; 
Imprints of $f (R, \Sigma, T)$ Gravity on early Universe via Big Bang Nucleosynthesis Observational Constraints. $J.~High~ Energy~ Astrophys.$ \textbf{53}, 100603 (2026).

\bibitem{mali} M. Ali, A. Eid, A. M. Sultan, M. U. Shahzad.; Big Bang Nucleosynthesis in Constrained $f(Q,C)$ Gravity: An Observational Analysis. $Forts.~ der~ Physik$ \textbf{74}, e70134 (2026).

\bibitem{47c} S. S. Mishra, A. Kolhatkar, P. K. Sahoo.; Big Bang Nucleosynthesis constraints on $f(T,\mathcal{T})$ gravity. $Phys. ~Lett.~ B$ \textbf{848}, 138391 (2024).

\bibitem{47d} N. S. Kavya, S. S. Mishra, P. K. Sahoo, V. Venkatesha.; Can teleparallel $f(T)$ models play a bridge between early and late-time Universe? $Monthly ~Notices~ of ~the ~Royal ~Astronomical ~Society$ \textbf{532}, 3126 (2024).

\bibitem{47e} A. Kolhatkar, S. S. Mishra, P. K. Sahoo.; Investigating early and late-time epochs in f(Q) gravity. $Eur.~ Phys.~ J.~ C$ \textbf{88}, 888 (2024).

\bibitem{47f} S. S. Mishra, S. Patel, P.K. Sahoo.; BBN to late-time acceleration in $f(T,\mathcal{L}_m)$ gravity. $Phys. ~Lett.~ B$ \textbf{872}, 140098 (2026).




\bibitem{52} U. Debnath.; Constructions of $f(R,G,T)$ 
 gravity from some expansions of the Universe $Int.~ J.~ Mod.~ Phys.~ A$ \textbf{35}, 2050203 (2020).

\bibitem{53} H. Chaudhary, A. Bouali, N.U. Molla, U. Debnath, G. Mustafa.; Cosmological tests of $f(R, G, T)$ dark energy model in FRW universe. $Eur. ~Phys.~ J.~ C$ \textbf{83}, 918 (2023).

\bibitem{53aa}  L. D. Landau and E. M. Lifshitz., The Classical Theory of Fields.  $Butterworth~Heinemann$, Oxford (1998).



\bibitem{54} J. Bernstein, L. S. Brown and G. Feinberg.; Cosmological Helium production simplified. $Rev~ Mod.~Phys.$  \textbf{61}, 25 (1989).
 
\bibitem{55} A .G. Cohen, A. Rujula and S. L. De Glashow.; A matter–antimatter universe? $ApJ$ \textbf{495}, 539 (1998).

\bibitem{59} D. F. Torres, H. Vucetich and A. Plastino.; Early universe test of non extensive statistics. $Phys.~Rev.~Lett.$ \textbf{79}, 1588 (1997).

 \bibitem{60} G. Lambiase.;  Dark matter relic abundance and big bang nucleosynthesis in Horava’s gravity. $Phys.~Rev.~D$ \textbf{83}, 107501 (2011).

 \bibitem{53a} F. M. Gonzalez et al.; An improved neutron lifetime measurement with UCN$_\tau$. $Phys.~Lett.~B$ \textbf{127}, 162501 (2021).
 
 \bibitem{62} G. Lambiase.: Lorentz invariance breakdown and constraints from big-bang nucleosynthesis. $Phys.~Rev.~D$ \textbf{72}, 087702 (2005).

 \bibitem{63} E. Aver and K. A. Olive and E. D. Skillman.; The effects of He I $\lambda10830$ on helium abundance determinations. $JCAP$ \textbf{07}, 011 (2015).

 \bibitem{64} R. J. Cooke and M. Pettini and C. C. Steidel.; One Percent Determination of the Primordial Deuterium Abundance. $Astrophys. ~J.$ \textbf{855}, 102 (2018).
 
\bibitem{65} Y. I. Izotov, T. X. Thuan.; The primordial abundance of $^4He$ revisited. $Astrophys.~J.$ \textbf{500}, 188 (1998).

 \bibitem{66} B. D. Fields and K.A. Olive.; On the evolution of helium in blue compact galaxies. $Astrophys.~J.$ \textbf{506}, 177 (1998).
 
 \bibitem{67} D. Kirkman et al.; The cosmological baryon density from the deuterium-to-hydrogen ratio in QSO absorption systems: D/H toward Q1243+3047. $Astrophys.~J.~Suppl.~Ser.$ \textbf{149}, 1 (2003).
 
\bibitem{68} Y. I. Izotov and T. X. Thuan.; Systematic effects and a new determination of the primordial abundance of $^4He$ and $dY/dZ$ from observations of blue compact galaxies. $Astrophys.~ J.$ \textbf{602}, 200 (2004).
 
\bibitem{69} C. Ranjit, P. Rudra, and S. Kundu.; Dynamical system analysis of modified chaplygin gas in Einstein-Aether gravity. $Eur.~Phys.~J.~Plus$ \textbf{129}, 208 (2014).

\bibitem{70} J. Alveya, N. Sabtib, M. Escuderoc and M. Fairbairnd.; Improved BBN constraints  on the variation of the gravitational constant. $Eur.~Phys.~ J.$ \textbf{80}, 148 (2023).

 
\bibitem{71} E. Di Valentino, et al.; The CosmoVerse white paper: addressing observational tensions in cosmology with systematics and fundamental physics. $Phys.~Dark~Universe$ \textbf{49} 10196 (2025). 

\bibitem{72} R. C. Nunes, S. Pan, E. N. Saridakis.; New observational constraints on $f(T)$ gravity from cosmic chronometers. $J.~Cosmol.~Astropart.~Phys.$ \textbf{1608}, 011 (2016).

\bibitem{73} R. Kou, C. Murray and J. G. Bartlett.; Constraining $f(R)$ gravity with cross-correlation of galaxies and cosmic microwave background lensing. $A~ \& ~A$ \textbf{686}, A193 (2024).

\bibitem{74} Y. C. Chen and C. Q. Geng and C. C. Lee and H. Yu.; Matter power spectra in viable $f(R)$ gravity models with dynamical background. $Eur.~Phys.~J.~C.$ \textbf{79}, 93 (2019). 
 
\end{thebibliography}
\end{document}